\documentstyle[12pt]{article}
\def\greaterthansquiggle{\raise.3ex\hbox{$>$\kern-.75em\lower1ex\hbox{$\sim$}}}
\def\lessthansquiggle{\raise.3ex\hbox{$<$\kern-.75em\lower1ex\hbox{$\sim$}}}
\newcommand{\beq}{\begin{equation}}
\newcommand{\eeq}{\end{equation}}
\newcommand{\beqa}{\begin{eqnarray}}
\newcommand{\eeqa}{\end{eqnarray}}
\newcommand{\beqan}{\begin{eqnarray*}}
\newcommand{\eeqan}{\end{eqnarray*}}
\newcommand{\ba}{\begin{array}}
\newcommand{\ea}{\end{array}}
\newcommand{\no}{\nonumber}

\newcommand{\Un}{\underline}
\newcommand{\ol}{\overline}

\newcommand{\D}{{\cal D}}

\newcommand{\cL}{{\cal L}}

\newcommand{\dsum}{\displaystyle \sum}

\def\nz{\ifmmode {I\hskip -3pt N} \else {\hbox {$I\hskip -3pt N$}}\fi}
\def\zz{\ifmmode {Z\hskip -4.8pt Z} \else
       {\hbox {$Z\hskip -4.8pt Z$}}\fi}
\def\qz{\ifmmode {Q\hskip -5.0pt\vrule height6.0pt depth 0pt
       \hskip 6pt} \else {\hbox
       {$Q\hskip -5.0pt\vrule height6.0pt depth 0pt\hskip 6pt$}}\fi}
\def\rz{\ifmmode {I\hskip -3pt R} \else {\hbox {$I\hskip -3pt R$}}\fi}
\def\cz{\ifmmode {C\hskip -4.8pt\vrule height5.8pt\hskip 6.3pt} \else
       {\hbox {$C\hskip -4.8pt\vrule height5.8pt\hskip 6.3pt$}}\fi}

\def\au{{\setbox0=\hbox{\lower1.36775ex%
\hbox{''}\kern-.05em}\dp0=.36775ex\hskip0pt\box0}}
\def\ao{{}\kern-.10em\hbox{``}}
\def\lint{\int\limits}
\normalsize
\begin{document}
\bibliographystyle{plain}

    \begin{titlepage}
    \begin{flushright}
    UWThPh--1998--6\\
    March, 1998
    \end{flushright}   
    \vspace{1.2cm}

    \begin{center}
    {\Large\bf  
    Dirac Variables and Zero Modes of Gauss Constraint \\[6pt]
    in Finite--Volume Two--Dimensional QED}\\

    \vspace{1.1cm}
    S. Gogilidze $^{a}$\\
    \medskip
    Institute for High Ehergy Physics\\
    University of Tbilisi\\
    Chavchavadze 3, 080 Tbilisi, Georgia\\
    \vspace{0.5cm}
    Nevena Ilieva $^{\ast, b}$\\
    \medskip
    Institute for Theoretical Physics\\
    University of Vienna\\
    Boltzmanngasse 5, 1090 Vienna, Austria\\
    \vspace{0.5cm}
    V.N.Pervushin $^{c}$\\
    \medskip
    Bogoliubov Laboratory of Theoretical Physics\\
    Joint Institute for Nuclear Research\\
    141980 Dubna, Moscow Region, Russia\\

    \vfill
    {\bf Abstract}\\
    \end{center}

    The finite--volume QED$_{1+1}$ is formulated in terms of Dirac
    variables by an explicit solution of the Gauss constraint with possible
    nontrivial boundary conditions taken into account. The intrinsic
    nontrivial topology of the gauge group is thus revealed together with
    its zero--mode residual dynamics. Topologically nontrivial gauge
    transformations generate collective excitations of the gauge
    field above Coleman's ground state, that are completely decoupled
    from local dynamics, the latter being equivalent to a
    free massive scalar field theory.

    \vspace{0.4cm}
    \vfill
    {\footnotesize

    $^\ast$  On leave from Institute for Nuclear Research and Nuclear Energy,
    Bulgarian Academy of Sciences, Bul.Tzarigradsko Chaussee 72, 1784 Sofia,
    Bulgaria

    $^a$ E--mail address: soso@jinr.ru

    $^b$ E--mail address: ilieva@pap.univie.ac.at

    $^c $ E--mail address: pervush@thsun1.jinr.ru}

    \end{titlepage}
    \section{Introduction}

    Separation of real dynamical variables from nondynamical ones
    is the crucial step in extracting the relevant physical
    information from the gauge theories. The usual method for achieving
this purpose --- by imposing a gauge--fixing condition, might not be always
adequate to the dynamical content of the classical equations of motion.
Another possibility, offered by Dirac \cite{Dir}, consists in introduction
of gauge invariant dynamical variables through an explicit solution of the
    Gauss equation. Such an explicit solution might contain some additional
    physical information which is implicitly lost by the gauge fixing.
    This can be seen even on the simplest example of the
    two--dimensional QED. It is our aim in the present paper to construct the
    Dirac variables for this model and to clarify the physical 
    consequences from such an approach.

    We first solve the classical equations for the free electromagnetic
    field in a finite volume to obtain the residual dynamics on
    the Gauss constraint. We show that this dynamics could not be reproduced
    by an arbitrary gauge choice.

    Then we examplify the physical consequences of the existence of such a
    residual dynamics by the Schwinger model.

    \section{Free Electromagnetic Field in a Finite Volume}

    Let us start with the free Abelian gauge field in a finite--volume
    two--dimensional space--time:
    \beqa
    S & = & \lint_{-T/2}^{T/2}dt \lint_{-R/2}^{R/2}dx \, {\cL_0}(t,x)  \no \\
    {\cL_0}(t,x) & = & -\frac{1}{4} F^{\mu\nu}(t,x)F_{\mu\nu}(t,x) \, ,
    \eeqa
    where $F_{\mu\nu}$ is the field--strenght tensor.

    From this action, equations of motion follow
    \beq
    \partial^\mu F_{\mu\nu}(t,x) = 0 \, ,
    \eeq
    the one for $\nu = 0$ being actually a constraint:
    \beq
    \partial_1\partial_0 A_1(t,x) = \partial_1^{\, 2} A_0(t,x)\, .
    \eeq
    This equation can be explicitly integrated, thus yielding the solution
    \beqa
    A_0(t,x) & = & \frac{1}{2} \left[A_0(t,R/2) + A_0(t,-R/2)\right] +
    \frac{x}{R} \left[A_0(t,R/2) - A_0(t,-R/2)\right] + \no \\
    & + & \lint_{-R/2}^{R/2}\D(x,y)\,\partial_y \dot A_1(t,y)\, dy \, ,
    \eeqa
    where $\D(x,y)$ is the Green function for D'Alembert equation with zero
    boundary conditions
    $$
    \D(x,y) = \D(y,x)
    $$
    \beq
    \D(\pm R/2,y) = 0, \qquad \partial_y\D(\pm R/2,y) = 0
    \eeq

    Explicitly,
    \beq
    \D(x,y) = \frac{\vert x - y \vert}{2} + \frac{x\,y}{R} - \frac{R}{4}
    \, ,
    \eeq
    so that
    \beq
    \partial_x\,\partial_y\,\D(x,y) = -\delta(x-y) + \frac{1}{R}\, .
    \eeq

    We restrict ourselves to the case when $A_0(t,x)$ satisfies symmetric
     boundary conditions in $x$
     \beq
     A_0(t,-R/2) = A_0(t,R/2)\, .
     \eeq
     However, the $U(1)$ gauge invariance of the action (1)

    \beq A_\mu(t,x) \to A_\mu^g(t,x) =
    A_\mu(t,x) + \frac{i}{e}\partial_\mu \lambda(t,x)
    \eeq
    with $\lambda(t,x)$ an arbitrary function, means that the same property
    takes place also for the gauge transformations parameter $\,\dot
    \lambda(t,x)$.
    Then, the $x$--derivative of $ A_0 $ (4) with
    eqs.(5), (7) taken into account becomes 
    \beqa
    \partial_x A_0(t,x) & = &
\lint_{-R/2}^{R/2}\partial_x\D(x,y)\,\partial_y \dot A_1(t,y)\, dy = \no 
    \\
    & = & \dot A_1(t,x)  -
    \frac{1}{R}\lint_{-R/2}^{R/2}\dot A_1(t,y) dy
   \eeqa

    Thus, for the (only one) field--stregth tensor component we get
    $$
    F_{01}(t,x)\Bigr \vert _{\frac{\delta S}{\delta A_0} = 0} =
    \frac{1}{R}\lint_{-R/2}^{R/2}\dot A_1(t,y) dy
    = \ol F_{01}(t)\, .
    $$

    It is convenient to introduce a new collective variable
    \beq
    N(t) = \frac{e}{2\pi} \lint_{-R/2}^{R/2}A_1(t,x) dx \, ,
    \eeq
    so that
    \beq
    \ol F_{01}(t) = \frac{1}{R}\left(\frac{2\pi}{e}\right)\dot N(t)
    \eeq

    In terms of $N(t)$ the free action (1) reads
    \beq
    S^T = \frac{1}{2R}\left(\frac{2\pi}{e}\right)^2 \lint_{-T/2}^{T/2}
    \dot N^2(t) dt
    \eeq

     The remaining dynamical equation --- eq.(2) for $\nu = 1$, reduces to
     $$
     \ddot N(t) = 0
     $$
     which also follows from action (13). The collective variable $\,
     N(t)\,$ (11)
    can be cosidered as a continuous generalization of the Pontryagin index
    $\,\nu$,
    $$
    \nu = \frac{e}{4\pi}\lint_{-T/2}^{T/2}dt\lint_{-R/2}^{R/2}dx \,
    \epsilon_{\mu\nu}\,F^{\mu\nu}(t,x) = \lint_{-T/2}^{T/2}dt \dot N(t) =
    N(T/2) - N(-T/2)\, .
    $$

    As a functional on $\,A(t,x)$, this variable is invariant under gauge
    transformations
    (9), supposed the boundary condition (8) is simply applied to the gauge
    function $\,\lambda(t,x)\,$, thus resulting in
    $$
    \dot \lambda\left(t,-R/2\right) = \dot \lambda\left(t, R/2\right) .
    $$

    \section{Quantum Dynamics of the Collective Variable}

    The action so obtained, eq.(13), is relativistically invariant
    and can be straightforward quantized:
    \beq
    [N, P] = i, \qquad P = \frac{1}{R}\left(\frac{2\pi}{e}\right)^2 \dot N
    \eeq

    Thus, the gaugeless reduction of the action for the free Abelian gauge field
    results in a residual (actually longitudinal) action, describing the
    collective motion of the latter. This superfluid behaviour gives a
    deep insight into the interplay between the local dynamics and the
    global symmetries of the theory though could hardly be recovered in an
    arbitrary gauge.

    The solution of corresponding Schr\"odinger equation,
    invariant under (9), is of the form of a plane wave with respect to
    $\, N\,$
    \beq
    \Psi = e^{-iP N[A]}, \qquad P = \frac{2\pi}{e}E \, .
    \eeq
    Such an unsatisfactory result originates in the formal treatment of the
    boundary conditions for the problem. As is easily seen, both the variable
    (11) and the action (13) have no consistent classical interpretation.
    Therefore, when gauge transformations are to be considered as
    transformations of the quantum theory, the boundary conditions should be
    imposed on the gauge phase $\, e^{i\lambda(t,x)}\,$. In such a way,
    for the case at hand one has
    \beq
    e^{i\lambda(R/2)} = e^{i\lambda(-R/2)}
    \eeq
    that is
    \beq
    \lambda(R/2) - \lambda(-R/2) = 2\pi n, \qquad n \in {\bf Z}
    \eeq

    Thus, a mapping of the (finite) space $R(1)$ onto the group manifold of
    $U(1)$ is defined, with a (possibly) nontrivial degree of mapping $\,
    n\,$. The gauge fields are so devided into topological classes,
    characterized by the value of $n$, their configuration space acquiring
    the topology of a ring. However, points with different $\, n \,$ are
    physically equivalent, so that the
    wave function of the gauge field in them may differ only by a phase factor
    \beq
    \Psi[N+1] = e^{i\theta}\Psi[N], \qquad \vert \theta \vert < \pi 
    \eeq

    In such a theory, a purely quantum phenomenon --- a field Josephson effect,
    takes place. Due to the inhomogeneity of the wave function phase, eq.(3.5),
    a nonvanishing electric field without any sources is present in the ground
    state \cite{IP,P}.

    Indeed, now the collective variable $\, N[A]\,$ is only covariant under
    (9),(17) --- so, under topologically nontrivial or large gauge
    transformations,
    \beq
    N[A^g(t,x)] = N[A(t,x)] + n
    \eeq
    and the spectrum of its conjugate momentum $\,P\,$ is essentially
    changed due to the boundary condition (18). It indicates the equivalence
    of the states
    \beq \langle P\vert N \rangle = e^{iP N} \quad {\rm and}
    \quad \langle P\vert N+n \rangle = e^{iP (N+n)} ,
    \eeq
    the true state thus
    representing a Bloch wave
    \beqa \langle P\vert N \rangle & = &
    \lim_{l\to\infty}\frac{1}{l}\dsum_{n=-l/2}^{l/2}
    e^{in\theta}\,e^{-iP (N+n)} = \no \\
    \bigskip
    & = & \left\{\matrix{
    e^{-iN(2\pi k+\theta)}, & P = 2\pi k+\theta \cr
    0, & P \not= 2\pi k+\theta, \cr
    } \right. \no \\
    \bigskip
    & & k \in {\bf Z}, \qquad  \vert \theta \vert < \pi
    \eeqa

    This means that the operator $\hat E$ describes a constant electric field
    and its spectrum is
    \beq
    \hat E \Psi = \frac{e}{2\pi}P\Psi = e(k + \frac{\theta}{2\pi}) \Psi
    \eeq
    So, we have obtained excitations above Coleman's ground state \cite{Col},
    which
    corresponds to the minimal (in modulus) eigenvalue of the operator $\hat E$
    (for $k=0$).
    However, the interpretation is essentially different: while in \cite{Col}
    the existence of $\theta$, as an additional parameter, is justified by the
    properties of the space $R(1)$, here it appears as a bridge between the
    global symmetries and the local dynamics of the theory. Moreover,
    consideration of $\theta$ as an additional (continuous) parameter
    implies already a non--separable Hilbert space due to the existence
    of an uncountable set of orthogonal vectors (corresponding to different
    values of $\theta$), with all complications that follow from this.

    It should be emphasized that this is an entirely quantum effect. The
    residual topological action (13), as already mentioned, has
    no consistent
    classical interpretation, since the region of ``validity" of the quantum
    theory is proportional to the volume $R$ of the space. Therefore, in
    consideration of an infinite--volume model, the infrared regularization
    should be removed at the very final stage and only the final finite--volume
    results have to be checked for surviving the thermodynamic limit. An
    illustration of the importance of this observation is offered by the charge
    screening in the Schwinger model \cite{IJMP}.

    \section{Hamiltonian dynamics 
    in the first order formalism}
    
    As has been mentioned in Sec.3, the global excitation of the gauge
    field could be easily overlooked by an unappropriate gauge choice.
    However, this is not necessarily so, as shows consideration of the
    free action (1) in the first order formalism: 
    \beq
    S_I[E, A^\mu] = \lint_{-T/2}^{T/2}dt \lint_{-R/2}^{R/2}dx
    \left[ E(\dot A_1 - \partial A_0) - (1/2) E^2 \right] \, .
    \eeq
    The canionical momenta define the primary constraints 
    $$
    \phi_1 = \pi_E = 0, \qquad  \phi_2 = \pi_1 - E = 0, \qquad
    \phi_3 = \pi_0 = 0\, .
    $$
    Thus the total Hamiltonian takes the form
    $$
    H_T (t) = \lint_{-R/2}^{R/2}dx \left[ (1/2) E^2 + E\partial A_0 +
    u_1 \pi_E + u_2 (\pi_1 - E) + u_3 \pi_0 \right] .
    $$

    Among the primary constraints there are one first class constraint and two
    second class ones
    $$
    \{ \phi_1, \phi_2 \} = \delta (x-x')
    $$
    $$
    \{ \phi_1, \phi_3 \} = \{ \phi_2, \phi_3 \} = 0
    $$
    The stationarity conditions for the primary constraints fix two Lagrange
    multipliers
    $$
    u_1 = 0 \qquad u_2 = E - \partial A_0
    $$
    and generate one secondary constraint
    $$
    \Phi = \partial E = 0 ,
    $$
    which is itself stationary so that no more constraints appear in the
    theory.

    With the following admissible gauge--fixing conditions:
    $$
    \chi_1 = A_0 = 0 \qquad
    \chi_2 = \pi_E = 0
    $$
    the final Hamiltonian is obtained
    $$
    H(t) = \lint_{-R/2}^{R/2}dx \left[ (1/2) E^2 + E\partial A_0 +
	   + ( E - \partial A_0)(\pi_1 - E) \right] .
    $$
    Thus the Hamiltonian equations of motion are obtained to be
    $$
    \dot A_0 = 0, \quad
    \dot \pi_0 = 0, \quad
    \dot A_1 = E - \partial A_0, \quad
    \dot \pi_1 = 0, \quad
    \dot E = 0, \quad
    \dot \pi_E = 0,
    $$
    $$
    \pi_E = 0, \quad
    \pi_1 - E = 0, \quad
    \pi_0 = 0, \quad
    \partial E = 0, \quad
    A_0 = 0\, . \quad
    $$
    The solution of this system is
    $$
    E(x,t) = \frac{e}{2\pi} P(t)
    $$
    and depending on the boundary conditions, reproduces our previous
    results. 

    \section{Dirac Observables for the Finite--Volume QED${\bf _{1+1}}$}

    The finite--volume Schwinger model is characterized by the action
    $$
    S = \lint_{-T/2}^{T/2}dt \lint_{-R/2}^{R/2}dx \, \cL(\Un x)
    $$
    with Lagrangean density
    \beq
    \cL(\Un x) = -\frac{1}{4} F^{\mu\nu}(\Un x)F_{\mu\nu}(\Un x) +
    i\bar\psi(\Un x) \gamma_\mu \left(\partial^\mu - ieA^\mu(\Un x)\right)
    \psi(\Un x),
    \eeq
    $$
    \Un x = (t,x), \qquad {\rm diag} \: g_{\mu\nu} = (1, -1)
    $$
    where $\psi(x), \bar\psi(x)$ are anticommuting two--component spinors
    and the two--dimensional $\gamma$--matrices are
    $$
    \gamma_0 = \sigma_1, \quad \gamma_1 = -i\sigma_2, \quad
    \gamma_5 = \gamma_0\gamma_1 = \sigma_3 \, ,
    $$
    with $\sigma_k$ being the Pauli matrices.

    Equations of motion now read
    \beqa
   & -\partial^\mu F_{\mu\nu}(\Un x) = J_\nu(\Un x), \qquad
    J_\mu(\Un x) = \bar\psi(\Un x)\gamma_\mu\psi(\Un x) \no \\
   & \gamma_\mu \left(\partial^\mu - ieA^\mu(\Un x)\right)
    \psi(\Un x) = 0
    \eeqa
    so that the constraint --- eq.(25a) for $\nu=0$, depends also on the
    current
    \beq
    \partial_1^{\,2} A_0(\Un x) = \partial_1 \partial_0 A_1(\Un x) + J_0(\Un x)
    \eeq

    An explicit solution of the constraint equation can still be obtained
    \beqa
    A_0(t,x) & = & \frac{1}{2} \left[A_0\left(t,R/2\right) +
    A_0\left(t,-R/2\right)\right] +
    \frac{x}{R} \left[A_0(t,R/2) - A_0(t,-R/2)\right] + \no \\
    & + & \lint_{-R/2}^{R/2}\D(x,y)\,\left[\partial_y \dot A_1(\Un y) +
    J_0(\Un y)\right]\,dy
    \eeqa
    with $\D(x,y)$ given by (5), (6).

   On the solution (27), the field--strenght tensor takes the form
    \beqa
    F_{01}(t,x)\Bigr \vert _{\frac{\delta S}{\delta A_0} = 0} & = &
    \frac{1}{R}\lint_{-R/2}^{R/2}\dot A_1(t,y) dy +
    \lint_{-R/2}^{R/2}\left(\frac{1}{2}\epsilon(y-x) - \frac{y}{R}\right)
    J_0(\Un y) dy \no \\
    & = & \ol F_{01}(t) - \ol Q +
    \frac{1}{2}\lint_{-R/2}^{R/2}\epsilon(y-x) J_0(\Un y) dy
    \eeqa
    with the charge $\ol Q$ defined as
    \beq
    \ol Q = \frac{1}{R}\lint_{-R/2}^{R/2}x J_0(\Un x) dx
    \eeq

    Following the procedure proposed by Dirac \cite{Dir}, we introduce now a
    gauge invariant set of variables --- the Dirac observables. This is achieved
    by a gauge transformation
    \beq
    A_\mu^D(\Un x) = A_\mu(\Un x) - \frac{\partial_\mu \lambda^D(\Un x)}{e},
    \qquad
    \psi^D(\Un x) = e^{-i\lambda^D(\Un x)} \psi(\Un x)
    \eeq
    with a gauge parameter
    \beq
    \dot\lambda^D(\Un x) = \frac{e}{2}\lint_{-R/2}^{R/2}\epsilon(x-y)
     \dot A_1(\Un y) dy + \frac{e}{2}\left[A_0(t,R/2) + A_0(t,-R/2)\right]
    \eeq
    which gives for the gauge field components
    \beqa
    A_1^D(\Un x) & = & 0 \\
    A_0^D(\Un x) & = & - x \,\ol F_{01}(t) +
    \lint_{-R/2}^{R/2} \D(x,y) J_0(\Un y) dy \no
    \eeqa

    With (28) in mind, the reduced Lagrangean can be written, entirely
    in terms of observables (30), (32):
    \beqa
    L &=& i\bar\psi^D {\gamma}_{\mu} \partial^\mu \psi^D +
    \frac{1}{2R}\left(\frac{2\pi}{e}\right)^2
    \dot N^2
    -  2\left(\frac{2\pi}{e}\right)\dot N \ol Q + \no \\
    &+& \frac{1}{2}\lint_{-R/2}^{R/2} J_0(\Un x)\D(x,y) J_0(\Un y) \,dx\,dy
    \eeqa
    Here, relation (12)  has also been used in order to get the
    collective variable $\,N[A]\,$ explicitly entering the reduced action,
    so that the interaction between the local fields and the global
    collective mode becomes apparent.

    \section{Currents, Symmetries and Charge Conservation in Finite Volume}

    As is well known, the current commutator in two dimensions is afflicted
    with an anomaly --- the so--called Schwinger term,
    \beq
    \left[ J_0(x), J_1(y)\right] = \frac{ie^2}{\pi} \delta'(x-y) \, ,
    \eeq
    which is just a manifestation of the Dirac vacuum needed for a consistent
    quantum theory to be written, even in the case of free fermions, but also
    when various types of interactions are present (see, for example
    \cite{J,ML,IP}, also \cite{G} for the 4--dimensional case).

    Relation (34) is the ground for the bosonization in two dimensions,
    which is formally achieved (for fermion bilinears) through the following
    identification
    \beq
    J_{5\mu}(x) = \frac{e}{\sqrt\pi}\partial_\mu \phi(x), \qquad
    J_{5\mu} = \bar\psi(x)\gamma_5\gamma_\mu\psi(x)
    \eeq

    Thus, from Lagrangean (33) the following reduced Hamiltonian is
    obtained
    \beqa
    H^{red} &=& \frac{1}{2}R\left(\frac{e}{2\pi}\right)^2
    \left(P_N + \frac{4\pi}{e}\ol Q\right)^2 - \frac{e^2}{2\pi}R\ol\phi^2 +
    \no \\ &+& \frac{1}{2}\lint_{-R/2}^{R/2}\left[ \Pi^2(x) +
    \left(\partial_1\phi(x)\right)^2 + \frac{e^2}{\pi}\phi^2(x)\right] dx
    \, ,
    \eeqa
    where $\,\Pi(x)\,$ is the field--$\phi$ conjugate momentum
    $$
    \left[ \phi(x), \Pi(y)\right] = i\delta(x-y), \qquad \Pi(x) \equiv
    \partial_0\phi(x) = \frac{\sqrt\pi}{e} J_{50}(x) \, ,
    $$
    $P_N\,$ is the topological momentum
    \beq
    P_N =\frac{\partial L}{\partial \dot N}=
    \frac{1}{R}\left(\frac{2\pi}{e}\right)^2 \dot N
    - \frac{4\pi}{e}\ol
    Q, \qquad\left[N, P_N\right] = i
    \eeq
    and the ``mean field" $\,\ol \phi\,$ is defined as
    $$
    \ol\phi = \frac{1}{R}\lint_{-R/2}^{R/2}\phi(x) dx \, .
    $$

    The second term in the r.h.s. of eq.(36) has its origin in the
    finiteness
    of the space, hence in the specific form of the Green function $\,\D(x,y)\,$
    (see eqs.(6), (7)). Clearly, it vanishes in the limit $R \to\infty$,
    but we
    have already emphasized that the level of Hamiltonian is an early one for
    taking this limit. In what follows, we will specially underline the effects
    that are brought to being through this term.

    In the model under consideration, despite the usual electric and chiral
    charges
    \beqa
    Q_0 &=& \lint_{-R/2}^{R/2}J_{50}(x) dx = -\frac{e}{\sqrt\pi}
    \lint_{-R/2}^{R/2}\partial_1\phi(x) = -\frac{e}{\sqrt\pi}\left[\phi(R/2) -
    \phi(-R/2)\right]\\
    Q_5 &=& \lint_{-R/2}^{R/2}J_{50 }(x) dx = \frac{e}{\sqrt\pi}
    \lint_{-R/2}^{R/2} \Pi(x) dx
    \eeqa
    also the charge $\ol Q$ (29) can be constructed, which is related to the
    field $\ol\phi$ through the equivalence relations (35)
    \beq
    \ol Q = \frac{e}{R\sqrt\pi}\lint_{-R/2}^{R/2}x\,\partial_1\phi(x) dx =
    \frac{e}{\sqrt\pi}\,\left[\,\ol\phi + (1/2)Q_0\,\right] \, .
    \eeq

    The zero--charge sector corresponds to a symmetric boundary condition on
    $\phi(x)$,
    \beq
    \phi(R/2) = \phi(-R/2) \, ,
    \eeq
    in this case the charge $\ol Q$ is simply proportional to $\ol\phi$ and
    vanishes together with it in the limit $R\to\infty$. At finite $R$,
    it does not correspond to an invariance of the theory
    $$
    e^{i\alpha\ol Q}\,H^{red}\,e^{-i\alpha\ol Q} = H^{red} - \alpha Q_5 /R
\, ,
    $$
    but this is also the case in the original fermionic formulation. The
    situation with the chiral charge, hence with the (global) chiral symmetry,
    is more subbtle: chiral transformations, implemented by $\,\sigma =
    exp\{i\sqrt\pi Q_5/e\}\,$, do correspond to
    an invariance of the original fermionic action (1). The
    descendent of the fermions --- field $\phi(x)$, is
    itself not invariant but instead is shifted by an integer
    $$
    e^{inQ_5\sqrt\pi/e}\,\phi(x)\,e^{-inQ_5\sqrt\pi/e} =
    \phi(x) + n , \qquad n \in {\bf Z}
    $$

    The dynamics, due to the interplay between this local field and the
    collective mode $\,N\,$, is nevertheless chiral invariant and the charge
    $Q_5$ is conserved
    \beq
    i\,[\,H^{red}, \,Q_5\,] = 0
    \eeq

    Thus, there is no global chiral symmetry broken to worry about, so no
    ground for an $U(1)$ problem. It is worthwhile
    noticing that (42) results from cancellation of contributions, generated
    by the boundary term and the genuine mass term, and by vanishing of the
    contribution from the kinetic term alone. This would not be the case if we
    would have removed the infrared regularization already in the reduced
    Hamiltonian, since then there would be nothing to compensate the mass--term
    contribution and the global chiral symmetry would have been broken.

    The local chiral current, however, remains anomalous,
    \beqa
    \partial^\mu J_{5\mu} &=&
    \left(\frac{e}{\sqrt\pi}\right)^3
    (\phi(x) - \ol\phi) - \no \\
    &-& \frac{e}{\sqrt\pi} \left[ \,\phi'(R/2)\,\delta(R/2-x) -
    \phi'(-R/2)\,\delta(R/2+x)\,\right]\, .
    \eeqa

    With the reasonable assumption
    $$
    \phi'(R/2) = \phi'(-R/2)
    $$
    and with (35) in mind,
    eq.(43) takes the form of the Klein--Gordon equation for a free
    massive scalar field
    \beq
    \left(\Box + m^2\right)\tilde\phi(x) = 0, \qquad m = e/\sqrt\pi
    \eeq
    This is not an equation for the field $\phi(x)$ itself, but for the field
    \beq
    \tilde\phi(x) = \phi(x) - \ol\phi
    \eeq

    This should be expected since the Hamiltonian (36) can be represented
also
    as
    \beqa
    H^{red} &=& \frac{1}{2}R\left(\frac{e}{2\pi}\right)^2
    P_N^{\,2} + 2R\ol Q\left(\ol Q + \frac{e}{2\pi}P_N\right) + \no
    \\
    &+& \frac{1}{2}\lint_{-R/2}^{R/2}\left[ \Pi^2(x) +
    \left(\partial_1\tilde\phi(x)\right)^2 +
    \frac{e^2}{\pi}\tilde\phi^2(x)\right] dx
    \eeqa
    which makes (44) apparent.

    There is one special case,
    \beq
    \ol\phi = -\frac{P_N}{2\sqrt\pi}
    \eeq
    so then $\tilde\phi(x)$ appears as a combined field
    \footnote{Still one has to think about the precise meaning of this field in
    the context of (21)}

    \beq
    \tilde\phi_s(x) = \phi(x) + \frac{P_N}{2\sqrt\pi}
    \eeq

    In this case there is no explicit interaction term, though the interaction
    between the global mode $N$ and  the local
    field $\tilde\phi(x)$ is intrinsic due
    to the very structure of the latter. Now the chiral charge is no longer
    conserved
    $$
    \dot Q_5 = m^3 R P_N/\sqrt\pi
    $$
    and the axial current, still being anomalous, does not provide a free
    equation for $\tilde\phi(x)$ but either an interacting one for the latter
    \beq
    \left(\Box + m^2\right) \tilde\phi(x) = m^2 \frac{p_N}{2\sqrt\pi}
    \eeq
    or a free equation but for the original field $\phi(x)$
    \beq
    \left(\Box + m^2\right) \left(\tilde\phi(x) + \bar\phi\right) = 0 \, .
    \eeq

    However, the action is invariant under simultaneous transformations of
both
    fields
    \beq
    \tilde\sigma^n\,H^{red}[\tilde\phi_s]\,\tilde\sigma^{-n} =
    H^{red}[\tilde\phi_s], \qquad \tilde\sigma = e^{i\sqrt\pi(Q_5-2eN)/e} =
    e^{i\sqrt\pi\tilde Q_5/e}
    \eeq
    which have to be considered as the new chiral transformations. The charge
    $\tilde Q_5$ is conserved but no local current can be associated with it.

    \section{Concluding Remarks}

 We have examplified by two simple models the advantages of the
application of Dirac variables for quantization of gauge theories.
Introduction of these variables based on the explicit solution of the
constraint equations, provides some additional dynamical information,
which is implicitly lost by the gauge fixing.

Thus,  already in the case of a free two--dimensional electromagnetic
field in a finite volume, the zero--mode sector shows a superfluid
behaviour, due to the residual topological dynamics of the Abelian gauge
field, which however survives in the thermodynamic limit. It is the
existence of this global (topological) mode that is responcible for the
appearance of the $\theta$--vacuum, which is therefore present also in
the free theory.

Consideration of the finite--volume Schwinger model in the same spirit
leads to an unexpected result, namely the conservation of the chiral
charge, though the local chiral current remains anomalous.  This is now
only a reason for the mass generation and not an explanation for the
absence of the Nambu--Goldstone phenomenon, since the latter has simply
no ground in this case. The $\theta$--vacuum structure is nevertheless
justified also in the finite--volume considerations. Thus, the model
examplifies the independence between the global chiral symmetry (its
breaking and restoration) and the nontrivial vacuum structure, underlying
the purely topological origin of the latter.

Though the global mode appears to be formally decoupled from local
dynamics, its contribution is crucial for the symmetry properties of the  
model. This shows once again the severe influence of the spurious
deggrees of freedom on the physical interpretation of the
quantized constraint dynamics.

    \section*{Acknowledgements}

    It is a pleasant duty to thank R. Bertlmann, V.A. Efremov, A.M.
    Khvedelidze, E.A.~Kuraev and W. Thirring for stimulating discussions.
    We acknowledge hospitality at the Institute for Theoretical Physics of
    Vienna University, where this paper has been completed.

    \medskip
    This work has been supported in part by the ``Fonds zur F\"orderung der
wissenschaftlichen Forschung in \"Osterreich", Project P11287--PHY, and
by the Russian Foundation for Basic Investigation, Grant No. 96--01--01223. 

     \end{document}